\documentclass[DIV=calc, paper=a4, fontsize=10pt, twocolumn]{scrartcl}	 % A4 paper and 11pt font size

\usepackage{lipsum} % Used for inserting dummy 'Lorem ipsum' text into the template
\usepackage[english]{babel} % English language/hyphenation
\usepackage[protrusion=true,expansion=true]{microtype} % Better typography
\usepackage{amsmath,amsfonts,amsthm} % Math packages
\usepackage[svgnames]{xcolor} % Enabling colors by their 'svgnames'
\usepackage[hang, small,labelfont=bf,up,textfont=it,up]{caption} % Custom captions under/above floats in tables or figures
\usepackage{booktabs} % Horizontal rules in tables
\usepackage{fix-cm}	 % Custom font sizes - used for the initial letter in the document

\usepackage{sectsty} % Enables custom section titles
\allsectionsfont{\usefont{OT1}{phv}{b}{n}} % Change the font of all section commands

\usepackage{fancyhdr} % Needed to define custom headers/footers
\usepackage{lastpage} % Used to determine the number of pages in the document (for "Page X of Total")

\usepackage[pdftex]{graphicx}

\usepackage{lettrine} % Package to accentuate the first letter of the text
\newcommand{\initial}[1]{ % Defines the command and style for the first letter
\lettrine[lines=3,lhang=0.3,nindent=0em]{
\color{CadetBlue}
{\textsf{#1}}}{}}

\usepackage{titling} % Allows custom title configuration

\newcommand{\HorRule}{\color{CadetBlue} \rule{\linewidth}{1pt}} % Defines the CadetBlue horizontal rule around the title

\pretitle{\vspace{-30pt} \begin{flushleft} \HorRule \fontsize{50}{50} \usefont{OT1}{phv}{b}{n} \color{DarkRed} \selectfont} % Horizontal rule before the title

\title{LArIAT: \\  Liquid Argon In A Testbeam} % Your article title

\posttitle{\par\end{flushleft}\vskip 0.5em} % Whitespace under the title

\preauthor{\begin{flushleft}\large \lineskip 0.5em \usefont{OT1}{phv}{b}{sl} \color{DarkRed}} % Author font configuration

\DeclareRobustCommand{\authorthing}{
\begin{tabular}{cc}
\texttt{F. Cavanna} & {\small Yale University}\\
\texttt{M. Kordosky} & {\small William \& Mary}\\
\texttt{J. Raaf, ~B. Rebel} & {\small FNAL}\\
\end{tabular}
}
\author{\authorthing}

\postauthor{ \usefont{OT1}{phv}{m}{} \color{Black}% Configuration for the institution name
~~~~~for the LArIAT Collaboration % Your institution

\par\end{flushleft}\HorRule}% Horizontal rule after the title

\date{June, 2014} % Add a date here if you would like one to appear underneath the title block

\begin{document}

\maketitle % Print the title

\thispagestyle{fancy} % Enabling the custom headers/footers for the first page 

{\Large \color{CadetBlue}
  \vspace{0.3in}
   {\sf \underline{The LArIAT Collaboration}}
  \vspace{0.1in}
}
{\sf\color{DarkRed}
  {}\\
  {}\\
  {\color{CadetBlue}\it Argonne (US)}: J.~Paley\\
  {\color{CadetBlue}\it Boston U. (US)}: D.~Gastler, E.~Kearns, R.~Linehan\\
  {\color{CadetBlue}\it Caltech (US)}: R.~Patterson\\
  {\color{CadetBlue}\it U. Chicago (US)}: W.~Foremen, J.~Ho, D.~Schmitz\\
  {\color{CadetBlue}\it U. Cincinnati (US)}: R.~Johnson, J.~St.John \\
  {\color{CadetBlue}\it Fermilab (US)}:  R.~Acciarri, P.~Adamson, M.~Backfish, W.~Badgett, B.~Baller, A.~Hahn, D.~Jensen, T.~Junk, M.~Kirby, T.~Kobilarcik, 
  P.~Kryczynski, H.~Lippincott, A.~Marchionni, K.~Nishikawa, J.~Raaf, E.~Ramberg, B.~Rebel, M.~Stancari, G.~Zeller \\
  {\color{CadetBlue}\it Imperial College London (UK)}: M.~Wascko, \\
  {\color{CadetBlue}\it KEK (Japan)}: T.~Maruyama, E.~Iwai, S.~Kunori \\
  {\color{CadetBlue}\it LANL (US)}: C.~Mauger\\
  {\color{CadetBlue}\it U. L'Aquila (Italy)}: F.~Cavanna (presently at {\it Yale U.})\\
  {\color{CadetBlue}\it LNGS-INFN (Italy)}: O.~Palamara (presently at {\it Yale U.})\\
  {\color{CadetBlue}\it Louisiana State U. (US)}: F.~Blaszczyk, W.~Metcalf, A.~Olivier, M.~Tzanov\\  
  {\color{CadetBlue}\it Manchester U. (UK)}: J.~Evans, P.~Guzowski\\
  {\color{CadetBlue}\it Michigan State U. (US)}: C.~Bromberg, D.~Edmunds, D.~Shooltz\\
  {\color{CadetBlue}\it U. Minnesota Duluth (US)}: R.~Gran, A.~Habig, K.~Kaess\\
  {\color{CadetBlue}\it U. Pittsburgh  (US)}: S.~Dytman\\
  {\color{CadetBlue}\it Syracuse U. (US)}:  J.~Asaadi, M.~Soderberg, J.~Esquivel\\ 
  {\color{CadetBlue}\it U. Texas Arlington (US)}:  A.~Farbin, S.~Park, J.~Yu\\
  {\color{CadetBlue}\it U. Texas Austin (US)}:  J.~Huang, K.~Lang \\  
  {\color{CadetBlue}\it U. College London (UK)}: R.~Nichol, A.~Holin, J.~Thomas \\  
  {\color{CadetBlue}\it William and Mary Coll. (US)}: M.~Kordosky, M~Stephens, P.~Vahle\\         
  {\color{CadetBlue}\it Yale U. (US)}: B.T.~Fleming, F.~Cavanna, E.~Church, E.~Gramellini, O.~Palamara, A.~Szelc\\
%   {}\\  
}

%----------------------------------------------------------------------------------------
%	ABSTRACT
%----------------------------------------------------------------------------------------

% The first character should be within \initial{}
\initial{L}\textbf{iquid Argon Time Projection Chambers (LArTPCs) are ideal detectors for precision neutrino physics.  These detectors, when located deep underground, can also be used for measurements of proton decay, and astrophysical neutrinos.  The technology must be completely developed, up to very large mass scales, and fully mastered to construct and operate these detectors for this physics program.  As part of an integrated plan of developing these detectors, accurate measurements in LArTPC of known particle species in the relevant energy ranges are now deemed as necessary.  The LArIAT program aims to directly achieve these goals by deploying LArTPC detectors in a dedicated calibration test beam line at Fermilab.  The set of measurements envisaged here are significant for both the short-baseline (SBN) and long-baseline (LBN) neutrino oscillation programs in the US, starting with MicroBooNE in the near term and with the adjoint near and far liquid argon detectors in the Booster beam line at Fermilab envisioned in the mid-term, and moving towards deep underground physics such as with the long-baseline neutrino facility (LBNF) in the longer term. }

%----------------------------------------------------------------------------------------
%	ARTICLE CONTENTS
%----------------------------------------------------------------------------------------
{\sf 
\section*{Introduction}

The LArTPC detector with its full 3D-imaging, excellent particle identification (PID) capability and precise calorimetric energy reconstruction represents the most advanced experimental technology for neutrino physics. The LArTPC technology is also recognized as a powerful tool in searches for proton decay especially in the K-neutrino decay mode. 

 %%%%%%%%%%%%%%%%%%% Figure 1 %%%%%%%%%%%%%%%%%%%%
\begin{figure}[!h]
\begin{centering}
%\vspace{-0.5cm}
\begin{tabular}{c}
\includegraphics[width=3.3in]{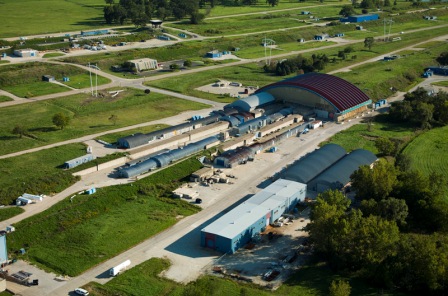}
\end{tabular}
%\vspace{-0.5cm}
\caption{
{\scriptsize \sf The Fermilab Test Beam Facility. Indicated by the arrow the MCenter beam line and the MC7 - LArIAT enclosure.}
}
\label{fig:ftbf}
\end{centering}
\end{figure}
%%%%%%%%%%%%%%%%%%%%%%%%%%%%%%%%%%%%%%%%%%

The ICARUS collaboration pioneered this technology in Europe \cite{Icarus} and demonstrated its feasibility for underground applications with an extended run at the INFN Gran Sasso Laboratory \cite{Icarus_at_GS}.  Interest in LArTPCs in the US has rapidly grown in recent years. First, the ArgoNeuT~\cite{ArgoNeuT} experiment collected neutrino data in the NuMI beam line at FNAL in 2009-2010 and is still producing physics results. This initial test experiment is being followed by MicroBooNE~\cite{MicroBooNE}, which is now in its final stages of construction, and is expected to be commissioned in 2014.  In 2012 LBNE~\cite{LBNE} made the LArTPC technology its choice for the far detector and advanced to CD-1 approval at the beginning of 2013. And finally, in 2014, a set of  near- \cite{LAr1ND}~and far- \cite{ICA-FNAL} ~liquid argon detectors, to be located in the Booster beam-line at FNAL together with MicroBooNE, have been proposed for sterile-$\nu$ search on short baseline.

Liquid argon detectors have been selected for these experiments because they promise excellent particle identification capabilities and calorimetric resolution, but only through experimental testing can these features be characterized in detail. Of particular interest are the features of electromagnetic shower reconstruction relevant to the separation of electron and gamma showers, the determination, without a magnetic field, of the sign of a muon via its endpoint capture or decay,  the identification of pions and kaons through the analysis of  their characteristic interaction modes in the liquid argon medium and the combined study of the ionization charge and scintillation photon yield to the deposited energy in the detector. 
 
 The LArIAT calibration test beam program in the Fermilab Test Beam Facility (FTBF)~\cite{FTBF} at Fermilab, shown in Fig.~\ref{fig:ftbf},  was proposed in 2012 and approved as T-1034 in 2013~\cite{LArIAT-MoU}. LArIAT is a DOE and NSF supported experiment, with additional substantial funds made available  through Fermilab's  LArTPC R\&D budget and Yale Univeristy through special agreement with the DOE.   A large international collaboration has been formed and progressively grown around this program. The collaboration currently includes more than 65 members from 17 institutions from the US, the UK, the EU and Japan. 

The experimental program is structured in two subsequent phases, with two LArTPC detectors of different sizes. We present here specifically the physics program and the detector layout for LArIAT Phase-1.

LArIAT Phase-1 will re-use the existing ArgoNeuT cryostat, which has a capacity of 550 liters, and its refurbished LArTPC, which contains 170 liters of active volume, equipped with new, signal-to-noise enhanced cold electronics derived from the MicroBooNE design. A system for the liquid argon scintillation light read-out, with augmented collection efficiency  is included in the detector layout.   A flexible trigger and data acquisition (DAQ) system based on commercially available devices is being developed for fast digitization of high rate test-beam interactions.  An over-sized cooling and filtering system is implemented to allow fast purification of the argon, stable and reliable operation during beam time and subsequent use with larger scale liquid argon volumes foreseen for Phase-2. Fermilab is providing engineering and technical support for the design and construction of both the DAQ and cryogenics systems. The LArIAT collaboration is responsible for the LArTPC, photo-sensors, trigger system, on-line monitoring and off-line data analysis. 

The dedicated beam line producing low momenta, $0.2 - 2.0$ GeV, particles is extracted from a high energy pion beam at the FTBF.  The implementation of this beam represents a combined effort from the FTBF and the LArIAT collaboration established  through the T-1034 TSW document~\cite{LArIAT-MoU} for the installation of the beam-line components including the target, collimators and bending magnets,  and of a series of beam detectors including wire-chambers, time-of-flight (TOF) counters, Cherenkov  counters and arrays of plastic scintillators necessary for beam characterization and trigger signal formation.

\section{Tertiary Beam Line}
\label{beam}
At the FTBF a primary beam consisting of 120~GeV$/c$ protons at moderate intensities from 1-300 kHz is targeted to create secondary charged particles in a wide range of energies for two beamlines, MTest and MCenter.
%%%%%%%%%%%%%%%%%%% Figure 2 %%%%%%%%%%%%%%%%%%%%
\begin{figure}[!h]
\begin{centering}
%\vspace{-0.5cm}
\begin{tabular}{c}
\includegraphics[height=2.3in]{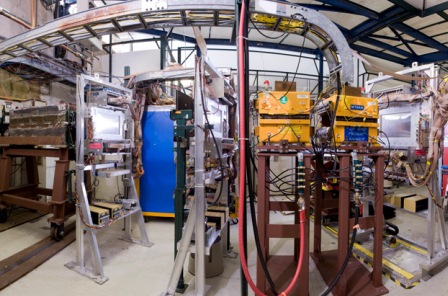}
\end{tabular}
%\vspace{-0.5cm}
\caption{
{\scriptsize \sf The tertiary beam components.}
}
\label{fig:tertiary}
\end{centering}
\end{figure}
%%%%%%%%%%%%%%%%%%%%%%%%%%%%%%%%%%%%%%%%%%
A four second spill duration, arriving every minute, is sent through the secondary beam line.  The MCenter line has been recently refurbished and has resumed operation. The line provides collimated pion beams in the 8 to 32 GeV range and has the added capability of a tertiary beam line, producing mainly pions and protons, but also at lower rate electrons, muons and kaons, of both signs, down to about 0.2 MeV/c. The LArIAT Phase-1 beam test is taking place along the MCenter line in the MC7 area, as shown in  Fig.~\ref{fig:ftbf}.

The tertiary beam consists of a target and collimator system and two bending magnets in a similar configuration used for the  MINERvA T-977 test beam calibration~\cite{MinervaTest}. A set of wire-chambers and TOF counters, also shown in Fig.~\ref{fig:tertiary}, provide tracking, momentum determination and particle identification (PID). The geometry of the tertiary beam line in MC7 has been optimized for LArIAT.  A 13$^\circ$ production angle through an upstream collimator and a 10$^\circ$ bending through a pair of dipole magnets are chosen, providing  particle momenta spectra tunable in the range from 0.2 to 2.0 GeV/c, as a function of the field intensity in the magnets. The schematic of the tertiary beam is shown in Fig.~\ref{fig:tert-layout} as rendered from the G4BeamLine MC simulation developed for beam configuration optimization.

%%%%%%%%%%%%%%%%%%% Figure 3 %%%%%%%%%%%%%%%%%%%%
\begin{figure}[!h]
\begin{centering}
\vspace{-0.3cm}
\includegraphics[height=1.3in]{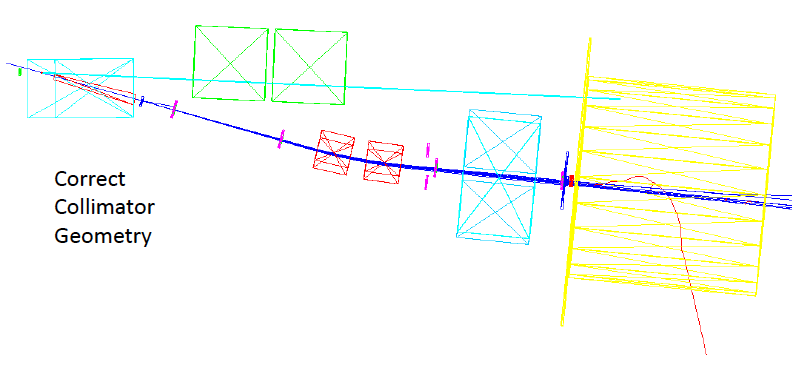}
\includegraphics[height=1.5in]{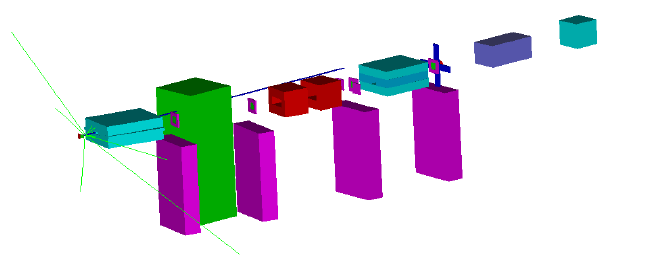}
\caption{
{\scriptsize \sf The tertiary beam layout from G4BeamLine MC simulation. Upstream and downstream collimators are in cyan, bending magnets in red, wire chambers and their stands in purple, the liquid argon TPC volume in violet.  In the top view, the cryostat is shown in yellow. The green block is for shielding.}
}
\label{fig:tert-layout}
\end{centering}
\end{figure}
%%%%%%%%%%%%%%%%%%%%%%%%%%%%%%%%%%%%%%%%%%
The LArIAT test experiment will run at MC7 for a period of at least two years, after beam operation start-up in 2014.

\section{Science}
\label{Science}

Calibration is a critical step to understanding the output response of any detector.  In particular, every new tracking detector or calorimeters is always calibrated before physics application.  

By ``calibration" we intend here the study in a LArTPC of all the particles emerging from neutrino interactions in the energy range relevant for the SBN and LBN programs by means of particles of species, momentum and sign emerging from the test beam in the desired energy range.  The test beam provides a controlled environment in which to develop and validate the off-line software tools for PID, calorimetry, and event reconstruction without relying solely on simulation. 

Figure~\ref{fig:mom-spectra} shows the momentum spectra of particles emerging from simulations  of neutrino interactions with the NuMI low-energy beam profile, which is quite similar to the planned LBNF.  The particle spectra are broadly distributed in the $0.2 - 2.0$ GeV$/c$ range and overlap the momentum spectra of the beam particles from the tertiary beam in MC7 quite well,  as shown in Fig.~\ref{fig:mom-spectra-beam}.
%%%%%%%%%%%%%%%%%%% Figure 4 %%%%%%%%%%%%%%%%%%%%
\begin{figure}[!h]
\begin{centering}
\vspace{-0.5cm}
%\begin{tabular}{c}
\includegraphics[height=2.5in]{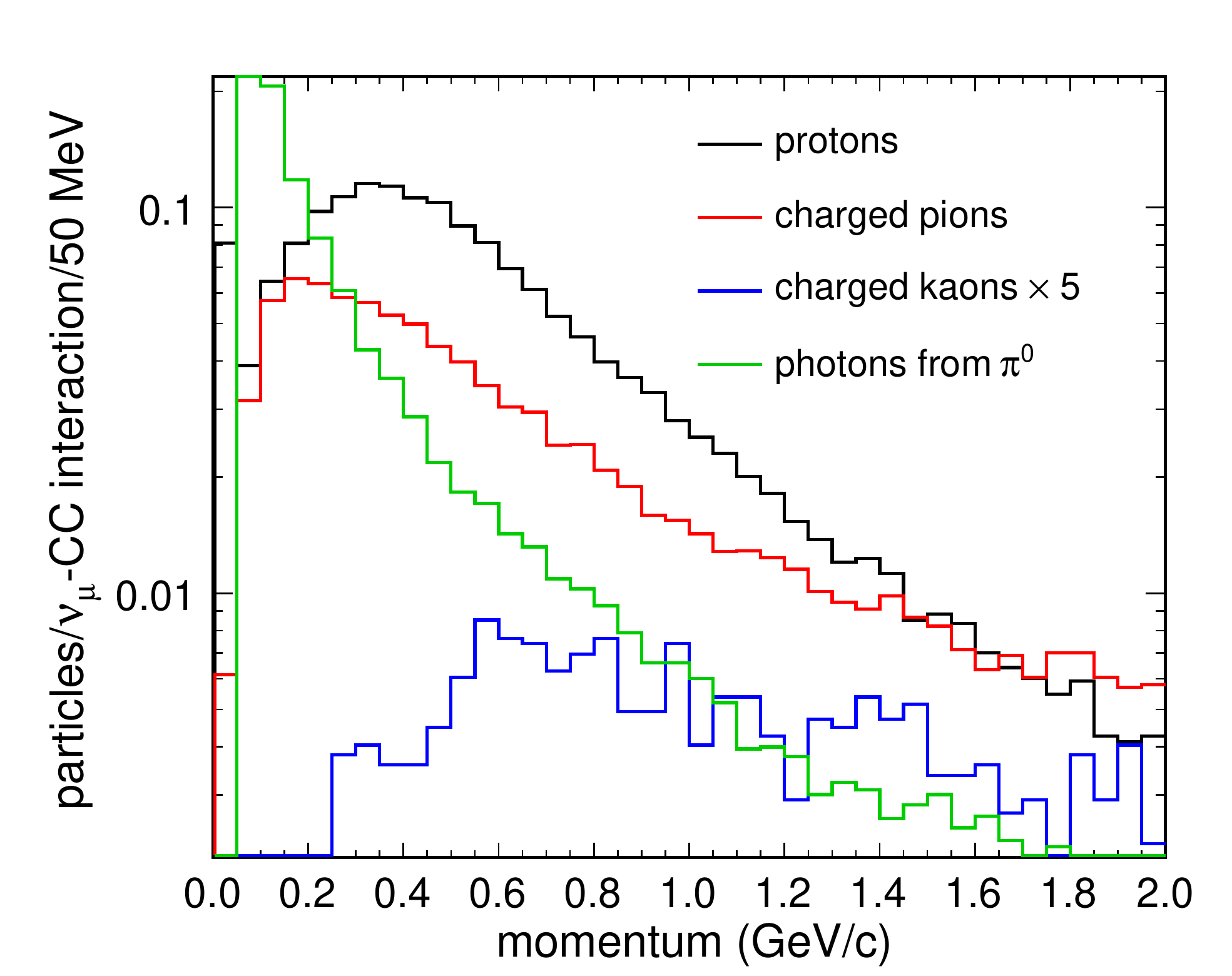}
%\end{tabular}
\vspace{-0.5cm}
\caption{
{\scriptsize \sf Momentum spectra of particles emerging from interaction of neutrinos with the NuMI LE beam profile}
}
\label{fig:mom-spectra}
\end{centering}
\end{figure}
%%%%%%%%%%%%%%%%%%%%%%%%%%%%%%%%%%%%%%%%%%

%%%%%%%%%%%%%%%%%%% Figure 5 %%%%%%%%%%%%%%%%%%%%
\begin{figure}[!h]
\begin{centering}
\vspace{-0.5cm}
%\begin{tabular}{c}
\includegraphics[height=2.5in]{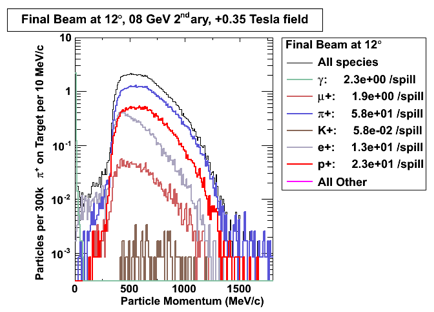}
\includegraphics[height=2.5in]{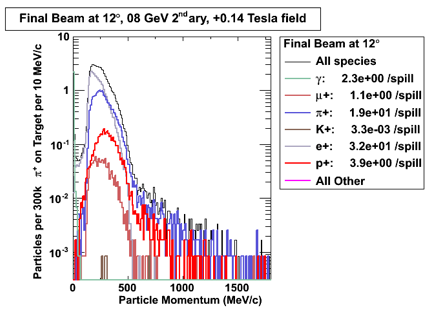}
%\end{tabular}
\vspace{-0.5cm}
\caption{
{\scriptsize \sf Momentum spectra of test beam particles produced with the LArIAT low energy tertiary beam, corresponding to different bending magnet settings.}
}
\label{fig:mom-spectra-beam}
\end{centering}
\end{figure}
%%%%%%%%%%%%%%%%%%%%%%%%%%%%%%%%%%%%%%%%%%

The LArIAT physics program is built over the availability of the test beam particles  from the tertiary beam in MC7.  The science goals of {\sf LArIAT} Phase-1 can be summarized as 
\begin{itemize}
\item[i)] experimental determination of the $e$ to $\gamma$-initiated shower separation;
\item[ii)] development of criteria for charge sign determination;
\item[iii)] single track calibration;
\item[iv)] optimization of pion and kaon identification through their interaction modes in argon medium;
\item[v)]  characterization of anti-proton stars in argon; and
\item[vi)] energy resolution improvement by combination of the scintillation light signal to the ionization charge.  
\end{itemize}
The goals are described in Sec.~\ref{e-gamma} to Sec.~\ref{antip}.

\subsection{ $e$-to-$\gamma$ shower separation}
\label{e-gamma}
The ability to separate electromagnetic showers from electrons compared to electromagnetic showers from photons is the key feature that led to the technology choice of LArTPCs for both the SBN and LBN detectors presently under construction or planned in the U.S.  The first category corresponds to the expected signal from charged-current interactions of electron neutrinos while the latter corresponds to the principal background from neutral pions produced in neutral-current interactions of all neutrino flavors.  The separation efficiency and sample purity for these two different categories of showers have never been experimentally measured, and current indications mainly rely on Monte-Carlo (MC) simulations. Only the initial part of the shower is relevant for separating these categories because the photon converts to an electron-positron pair producing double ionization in the first portion of the track where the two particles overlap at the shower start.  An experimental test can be performed using a small volume LArTPC.  Characterization with electron and bremsstrahlung photon beams, as shown in  Fig.~\ref{fig:e-gamma-sep}, will provide experimental confirmation for the separation efficiencies.  Such measurements will also further strengthen the physics case for both MicroBooNE as it attempts to identify of the low-energy $\nu_e$ excess observed by MiniBooNE,  and LBNE  as it measures the CP violating phase from the appearance of electron (anti-)neutrinos in a beam of muon (anti-)neutrinos.  Data from the test beam will readily enable more reliable separation algorithms to be developed and implemented in the LArSoft offline reconstruction code~\cite{larsoft}, thus benefitting multiple LArTPC experiments.
%%%%%%%%%%%%%%%%%%% Figure 6 %%%%%%%%%%%%%%%%%%%%
\begin{figure}[!h]
\begin{centering}
%\hspace{-0.3cm}
\begin{tabular}{c}
\includegraphics[height=1.3in]{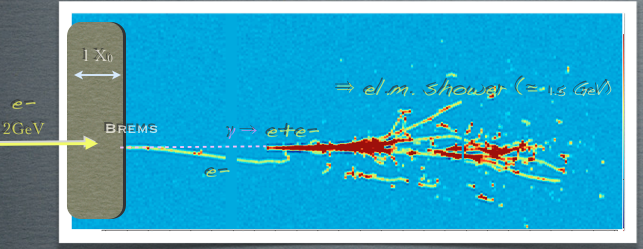}
\end{tabular}
\caption{
{\scriptsize \sf MC simaluation of a photon-shower induced by electron-bremsstrahlung in the LArIAT LArTPC.  The parent electron enters from the left. The photon conversion into a electron-positron pair initiates a shower that is well separated in the LArTPC volume from the parent electron track and shower.}
}
\label{fig:e-gamma-sep}
\end{centering}
\end{figure}
%%%%%%%%%%%%%%%%%%%%%%%%%%%%%%%%%%%%%%%%%%
 
\subsection {Charge sign determination}

The sign of a particle's charge can be obtained for stopping particles in LArTPC by statistical analysis based on topological criteria, even without a magnetic field. For example, $\mu^+$ undergo decay only, with $e^+$ emission of known energy spectrum. Stopping $\mu^-$ may either decay or be captured by nuclei. In Ar, the capture probability is about 76\%, accompanied by neutron and $\gamma$ emission.  The $\mu^-$-capture can thus be topologically separated from $\mu^+$-decay through the detection of a delayed Michel electron track in the LArTPC, as seen in Fig.~\ref{fig:mu-capt-mu-dec}.  The use of scintillation light can also contribute to the separation level \cite{Sorel}.  However, systematic study of the processes following $\mu^-$-capture in liquid argon have never been performed and the LArTPC sign determination capability has never been explored.  Beams with selectable polarity will provide data for direct measurement of the sign separation efficiency and purity for muons as well as for pions, and potentially also for kaons.  Capture topology and identification of the decay or capture products will further constrain the ability to distinguish the charge of the primary lepton in muon neutrino CC interactions of particular interest for CP violation, and for validating the reaction models implemented for argon nuclei in the GEANT4 simulation package \cite{geant4}. 
 %%%%%%%%%%%%%%%%%%% Figure 7 %%%%%%%%%%%%%%%%%%%%
\begin{figure}[!h]
\begin{centering}
%\vspace{-0.5cm}
\begin{tabular}{c}
\includegraphics[height=2.2in]{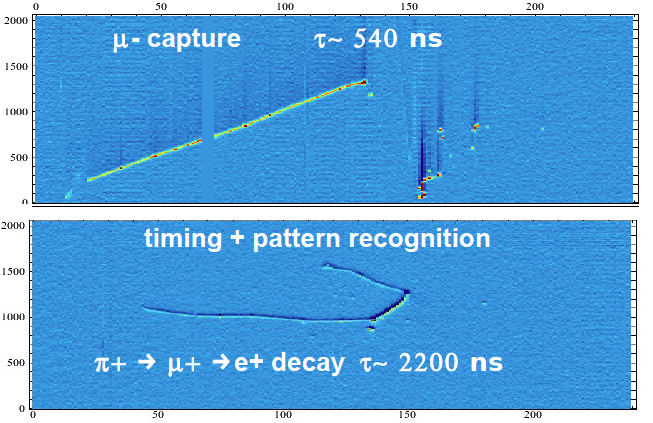}
\end{tabular}
\vspace{-0.5cm}
\caption{
{\scriptsize \sf Real events from ArgoNeuT showing examples of capture-like topology and decay-like topology. }
}
\label{fig:mu-capt-mu-dec}
\end{centering}
\end{figure}
%%%%%%%%%%%%%%%%%%%%%%%%%%%%%%%%%%%%%%%%%%

\subsection{ Single track calibration}
 
For liquid argon detectors the term calibration is often associated with the determination of the relationship between the collected ionization charge and the energy deposited in liquid argon by incident particles of different stopping powers. More specifically, calibration corresponds to performing measurements of electron-ion recombination in argon~\cite{miyajima,jaffe,onsager,box} over the most possible extended range of energy deposition rates, dE/dx; for different electric field values in the typical $0.3-1.0$~kV/cm range for LArTPC operation; and at different angles of incidence for the track with respect to the electric field. Augmenting the present knowledge of the fundamental properties of liquid argon as an active detection medium ultimately corresponds to enhance the calorimetric energy resolution of the detector. In particular, for LArTPCs the calorimetric energy reconstruction is possible at individual track level.
  
Pure low momentum beams of muons, pions, kaons and protons that penetrate and slow down to stop in the TPC would allow for an unprecedented, accurate calibration campaign. A modest volume LArTPC is sufficient for this task.
  
The dependence of the recombination on the electric field magnitude and direction in noble liquids is not fully understood, especially for high ionization densities~\cite{sala};  the data are fit with semi-empirical models~\cite{birks}.  Early data with energies above $15-20$ MeV/cm from the ICARUS measurements~\cite{fpp} were sparse and statistically limited.  Electron recombination in highly ionizing stopping protons and deuterons was more recently studied in the ArgoNeuT detector at a constant electric field\cite{argoneut_recomb}. The data are well-modeled by either a Birks model or a modified form of the Box model. However, the dependence of recombination on the track angle with respect to the electric field direction was found to be much weaker than the predictions of the Jaffe columnar theory and by simulations. 
  
The origin of this behavior represents an interesting topic for further investigation. Available beams of muons, pions, kaons and protons allow the collection of data with high statistics in different ranges of stopping power. Considering a minimum track pitch of 0.4 cm corresponding to the the LArIAT TPC wire pitch, the accessible stopping power range for muons (pions) goes from the mean value of a minimally ionizing particle of $dE/dx|_{mip}\simeq 2.1$  MeV/cm,  up to about 15~MeV/cm.  The value is extended to about 20~MeV/cm with kaons, and up to 30~MeV/cm with protons, that is $\sim 15\times dE/dx|_{mip}$, as seen in Table~\ref{tab:TB}.
 %%%%%%%%%%% Tab.1 %%%%%%%%%%%%%%%%%%%%%%%%%%%%%%
\begin{table}[h]
\centering
%\vspace{1cm}
\begin{tabular}{|c|c|c|c|}
	\hline\hline
              &   Incident            &  Incident       &  \\ 	
 Species&   Kinetic Energy  & Momentum  & $dE/dx$       \\ 	
              &   (MeV)               & (MeV$/c$)    & (MeV/cm)       \\ 	\hline
$\mu$    & 205              & 290   & $2.1\div13$  \\ 
$\pi$      & 210             & 320    & $2.1\div14$  \\ 
$K$        & 285            & 600    &  $2.8\div23$ \\ 
$p$        & 380            & 940     & $3.1\div33$ \\ 
\hline\hline
\end{tabular}
\caption{ {\scriptsize \sf We assume a contained-track maximum length for a stopping particle in liquid argon of $\ell_{max}= 80$ cm, which is less than the LArTPC longitudinal extension of 90~cm in LArIAT's TPC. Different charged particles correspond to different mean deposited energies which is equivalent to the incident kinetic energy of the particles, different incident beam momenta, and different intervals of mean dE/dx values along the track.}  }
\label{tab:TB}
\end{table}
%%%%%%%%%%%%%%%%%%%%%%%%%%%%%%%%%%%%%%%%%%%%%
The high accuracy and statistical precision of the test beam data will be fundamental to achieving an in-depth understanding of the recombination mechanisms in liquid argon.  It will also lead to an optimal model for their effects within the LArSoft off-line reconstruction and detector simulation package.
 
\subsection {Particle Identification}
When charged hadrons propagate through liquid argon and come to stop without inelastic interaction within the LArTPC volume, the energy deposited along the fully contained track can be combined with the range information from three dimensional tracking to compare the kinetic energy to the total range of the track as well as to compare the dE/dx to the residual range of the track, as seen in Fig.~\ref{fig:en-range-corr}. Efficient particle identification can be obtained using these measurements and this technique represents one of the key features of the LArTPC technology that is relevant for both neutrino oscillation experiments and proton decay searches. The separation of protons from pions and muons has been firmly established using ArgoNeuT data~\cite{argoneut_recomb}.  
  %%%%%%%%%%%%%%%%%%% Figure 8 %%%%%%%%%%%%%%%%%%%%
\begin{figure}[!h]
\begin{centering}
\includegraphics[height=3.0in]{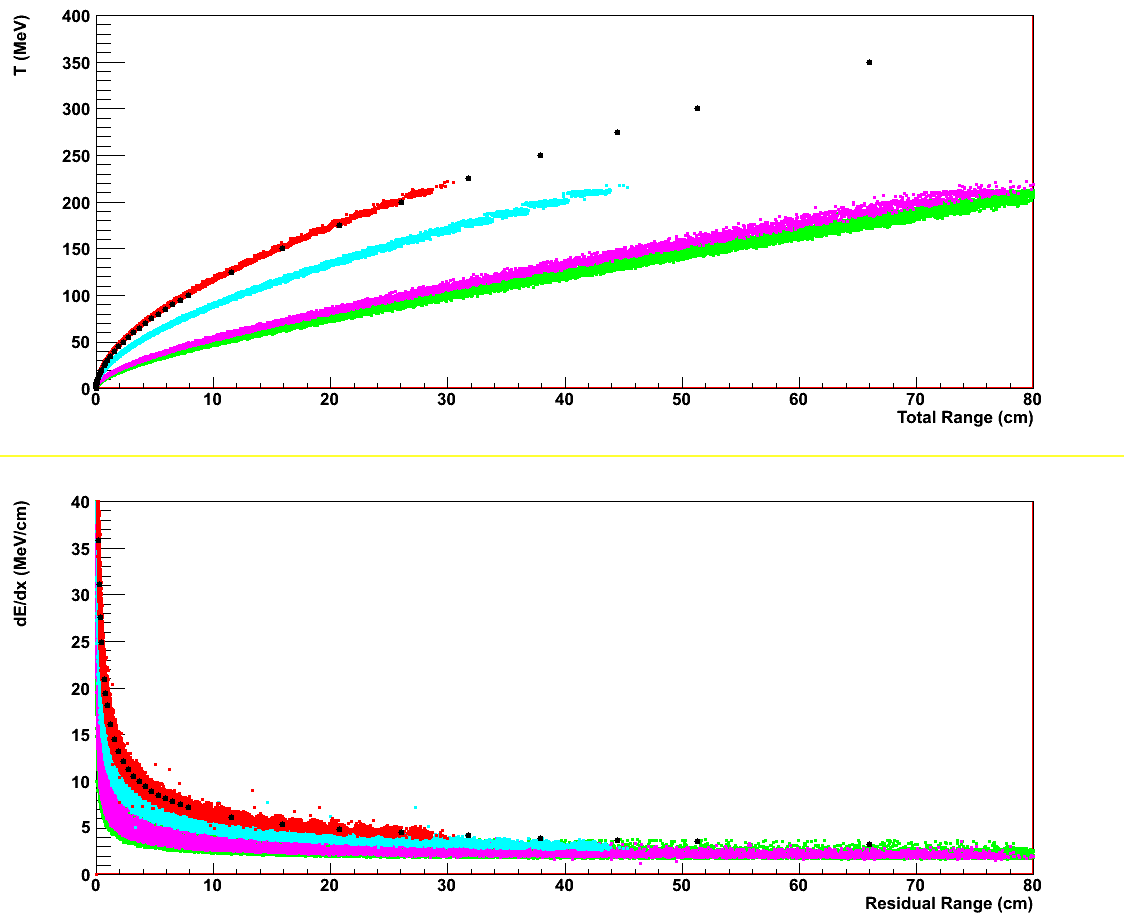}
\vspace{-0.5cm}
\caption{
{\scriptsize \sf  The deposited kinetic energy as a function of the total track range (top) and the dE/dx as a function of the residual track range (bottom) for $\mu$(green), $\pi$(violet), $k$(cyan) and $p$(red). }
}
\label{fig:en-range-corr}
\end{centering}
\end{figure}
%%%%%%%%%%%%%%%%%%%%%%%%%%%%%%%%%%%%%%%%%%

High statistics test beam data will allow experimental determination of proton to kaon identification efficiency and purity and 
kaon to $\pi/\mu$ identification efficiency and purity. This PID information, based on direct measurement with beam particles of known type, will greatly enhance confidence in the estimate of signal to background separation for neutrino cross-section studies with LArTPC detectors, such as ArgoNeuT, MicroBooNE, and LAr1-ND, as well as for future nucleon decay searches.

\subsection{Pion and Kaon interactions in Argon}
%%%%%%%%%%%%%%%%%%% Figure 9 %%%%%%%%%%%%%%%%%%%%
\begin{figure}[!h]
\begin{centering}
\includegraphics[height=1.6in]{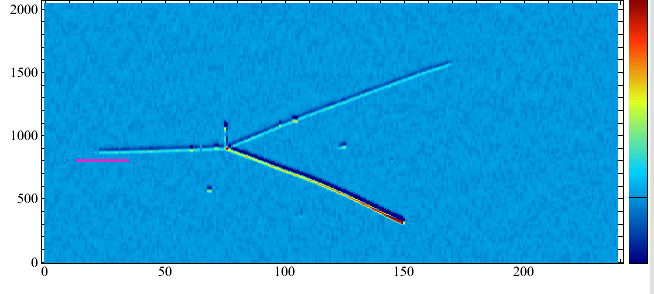}
%\end{tabular}
\vspace{-0.5cm}
\caption{
{\scriptsize \sf MC event with a $\pi^+$ undergoing inelastic interaction in Ar with proton knock-out (highly ionizing stopping track. }
}
\label{fig:piplus}
\end{centering}
\end{figure}
%%%%%%%%%%%%%%%%%%%%%%%%%%%%%%%%%%%%%%%%%%

One of the major sources of systematic uncertainty in precision neutrino oscillation measurements is the uncertainty on pion-nucleus interaction cross sections \cite{t2k}. The uncertainty is present at two levels: within the target nucleus of the primary neutrino interaction, and at the detector level with secondary interactions of pions that emerge from the target nucleus. Pion-nucleus interaction has four components: elastic scattering with the nucleus left in ground state, inelastic scattering with the nucleus excited or nucleon knock-out occurring,  absorption where no pion is present in the final state, and single or double charge exchange where a charged pion converts into either a neutral or oppositely charged pion.  For pions in the intermediate energy range of 100-500 MeV, as those of interest for LBNE and seen in Fig.~\ref{fig:mom-spectra}.  The pion interaction is dominated by the very strong $\Delta$(1232) resonance and the cross section is large. For example, at the 200 MeV $\Delta$ peak, the largest contribution to the pion cross section on the argon nucleus is expected to be from the absorption channel, with a cross section estimated to be $\sim$500 mb, extrapolated from measurements on lighter nuclei. 

The uncertainty, included in simulation codes such as GEANT for propagation in the detector, and GENIE for transport in  nuclear matter, is in the 30\% range or more depending on the channel. The goal of a dedicated pion run with LArIAT is initially to develop pion identification algorithms based on their interaction modes in argon. Ongoing studies will show the feasibility of a precise measurement of the pion-nucleus absorption  and charge exchange cross sections such to reduce the uncertainty on the neutrino interaction simulations.

\subsection {Anti-proton stars in Ar}
\label{antip}
Low momentum $\bar p$ may allow the first study of hadron star topology from $\bar p p$ annihilation at rest in argon through the $\bar p$-Ar reaction (MC example in Fig.\ref{fig:anti-p}).
  %%%%%%%%%%%%%%%%%%% Figure 10 %%%%%%%%%%%%%%%%%%%%
\begin{figure}[!h]
\begin{centering}
\includegraphics[height=2.3in]{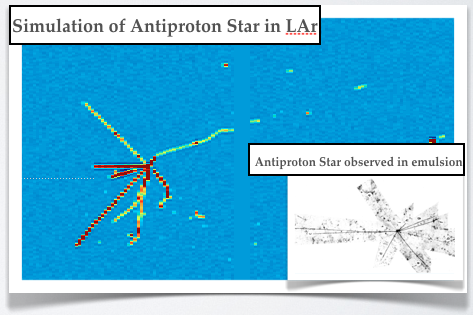}
%\end{tabular}
\vspace{-0.5cm}
\caption{
{\scriptsize \sf MC simulation of a $\bar pp$ annihilation at rest in argon. In the inset, the picture of a  $\bar p$ start in emulsions. }
}
\label{fig:anti-p}
\end{centering}
\end{figure}
%%%%%%%%%%%%%%%%%%%%%%%%%%%%%%%%%%%%%%%%%%
The multiplicity of $\pi^{\pm},\pi^0,K^{\pm},K^{L,S}$ in hadron stars can be accurately determined with a liquid argon imaging detector.  This information is considered very relevant for $n\bar{n}$-oscillation searches with future large underground LArTPC detectors.

Models for simulation of anti-proton annihilation at rest are the subject of continuous development within GEANT4~\cite{antip-annih}. Validation from experimental data is of interest as model predictions vary widely in the multiplicity and energy spectra for the secondaries produced.

In summary, the test beam run with LArIAT will provide vast amounts of useful data for a complete understanding of the fundamental mechanisms of energy release in a liquid argon target, immediately translating into calorimetric energy resolution improvement and enhancement of particle identification capability of LArTPC's.  As test beam data are accumulated, the LArSoft package will be upgraded through the optimization and development of data-based algorithms and methods for both off-line analysis and detector response simulation, providing a truly state-of-the-art software package for all LArTPCs. The GEANT4 modules for annihilation and capture at rest simulation of $\bar p$ and $\pi,K$ will also benefit from comparisons with experimental data.

The first experiments to benefit from the test beam results are the present generation LArTPC experiments, ArgoNeuT and  MicroBooNE.  All their relevant physics topics will be improved from higher confidence in the analysis algorithms, from exclusive channel cross-section measurements where robust and reliable PID is a necessary requisite for accurate neutrino interaction products recognition, to sterile neutrino searches with the best signal charged-current $\nu_e$ separation from neutral current $\pi^0$ background.  LBNE will also profit from these advances as it moves forward in its planning and construction. The LArIAT program will also train young physicists during extended beam operation and provide them with real data to analyze, which is also an invaluable contribution to the future SBN and LBN programs.

\section{Experimental Design: the {\sf LArIAT} Detector for Phase-1}
\label{sec:detector}
Any detector exposed to test beams must reproduce as closely as possible all technical features of the final detector.
%%%%%%%%%%% Figure 11 %%%%%%%%%%%%%%%%%%%%%%%%%%%%%%
\begin{figure}[h]
\begin{centering}
 \vspace{-0.4cm}
 \includegraphics[height=2.6in]{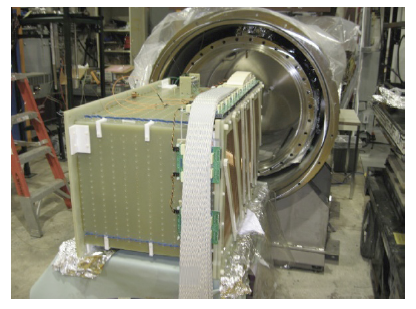}  
 \caption{ {\scriptsize \sf The fully instrumented ArgoNeuT TPC being inserted into the {\sf ArgoNeuT} cryostat opened through the removable end-cap.}  }
\label{fig:ArgoNeuT-pictures}
\end{centering}
\end{figure}
%%%%%%%%%%%%%%%%%%%%%%%%%%%%%%%%%%%%%%%%%%%%
Therefore, the single phase LArTPC scheme, with multiple wire-planes read-out geometry for 3D imaging, fine wire spacing and low-noise cold electronics, is the necessary choice for best informing the design and exploitation of all liquid argon detectors included in the present U.S. neutrino program\footnote{To date, there has been only one other set of measurements \cite{t32} taken using a prototype detector with coarse read-out sampling exposed to a test beam at KEK. This proposal expands on that work with a fine-grained LArTPC detector and a dedicated multi-year program.}.

The LArIAT Phase-1 experimental program at FTBF capitalizes on the availability of the existing hardware from the ArgoNeuT experiment.  The layout of the LArIAT experiment includes two main parts: the liquid argon-related components and the beam-related components. The liquid argon-related part consists of the LArTPC detector, the liquid argon scintillation light detector, the LArTPC read-out cold electronics, the liquid argon cryostat, and the cryogenic system connected to the cryostat for liquid argon cooling and purification. The beam-related part consists of a series of beam counters such as the TOF detector, multi-wire proportional chambers (MWPCs), Cherenkov detector, and veto paddles are aligned along the {\sf LArIAT} beam line for PID tagging and momentum selection.

The vacuum-insulated cryostat and the inner detector TPC are pre-existing components from ArgoNeuT and are shown in Fig.~\ref{fig:ArgoNeuT-pictures}. The active volume of the ArgoNeuT TPC is appropriate for the proposed {\sf LArIAT} Phase-1 physics program.  A detailed description of these components, cryostat and TPC, is available from the ArgoNeuT collaboration~\cite{tech_paper}, while the  modifications applied to these components for test beam operation and the new components of the LArIAT experimental set-up are outlined in the following sections. We start with the cryostat and then we proceed from the liquid argon detectors inside the cryostat towards the external systems, including the series of beam detectors  along the beam-line.

\subsection{The LArIAT Cryostat}
Liquid argon is contained in a stainless-steel, vacuum-jacketed and super-insulated cryostat, that was originally custom designed for ArgoNeuT.  The cryostat shape is cylindrical with convex end-caps. The main axis of the cryostat is horizontal and oriented parallel to the beam.  The inner vessel is $76.2$ cm (30'') in diameter ($\o$) and 130 cm in length corresponding to a liquid argon volume of about 550 L, or a mass of 0.76~t.  Access to the internal volume for detector installation is possible by opening the end-caps of the inner and outer vessels at one end of the cryostat.  The cryostat has a wide neck, or ``chimney",  located on the top of the cryostat at its mid-length and serves as access path for signal cables from the LArTPC and from the internal instrumentation, as well as for the high voltage (HV) feed-through. A series of radiation baffles designed to reduce the effects of cryogenic radiation are also placed inside the chimney.
%%%%%%%%%%% Figure 12 %%%%%%%%%%%%%%%%%%%%%%%%%%%%%%
\begin{figure}[h!]
\centering
\includegraphics[width=3.5in]{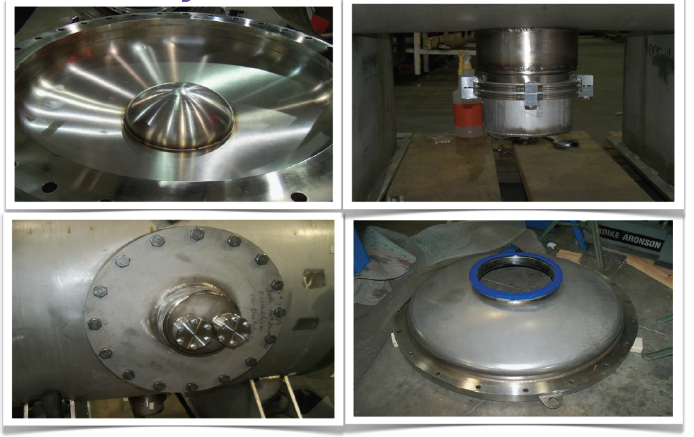}
\caption{ {\scriptsize \sf  Modifications to the cryostat including the excluder (top left), liquid argon drain outlet (top right), side ports for PMT read-out (bottom left), and the beam window (bottom right). } }
\label{fig:CryostatMods}
\end{figure}
%%%%%%%%%%%%%%%%%%%%%%%%%%%%%%%%%%%%%%%%%%%%

A series of hardware modification to the cryostat have been implemented to cope with the requirements of its use in a  charged particle test beam run.  These modifications include $i)$ a ``beam window" to reduce the amount of material upstream of the TPC active volume along the beam line, $ii)$ a connection for the argon cooling and purification system, and $iii)$ a connection for the scintillation light readout system.  In the ArgoNeuT cryostat,  the stainless steel front-ports of the cryostat inner and outer vessel were 3/16" (4.8 mm) thick and there was also a dead liquid argon layer of about 15~cm before reaching the TPC active volume inside the cryostat, for a total amount of about 1.6~$X_0$.   The 9-inch diameter beam window is located at the center of the outer vessel's front flange and is blanked with a 0.4 mm titanium sheet. There is also an excluder, or concavity,  in the front flange of the inner vessel of the cryostat to reduce the amount of dead liquid argon through which particles must travel before reaching the active volume. The total material thickness is now reduced to less than 0.3~$X_0$. Both cryostat vessels were modified to connect a liquid argon transfer line to the new cooling and purification system. The modification to the cryostat consists of an outlet at the bottom of the cryostat with Conflat and ISO flange sealing.  Both cryostat vessels had their side-ports modified with CF-flanged apertures for the signal and HV feedthroughs of the scintillation light system.

\subsection{LArTPC}
%%%%%%%%%%% Figure 13 %%%%%%%%%%%%%%%%%%%%%%%%%%%%%%
\begin{figure}[h!]
\begin{centering}
\includegraphics[width=3.5in]{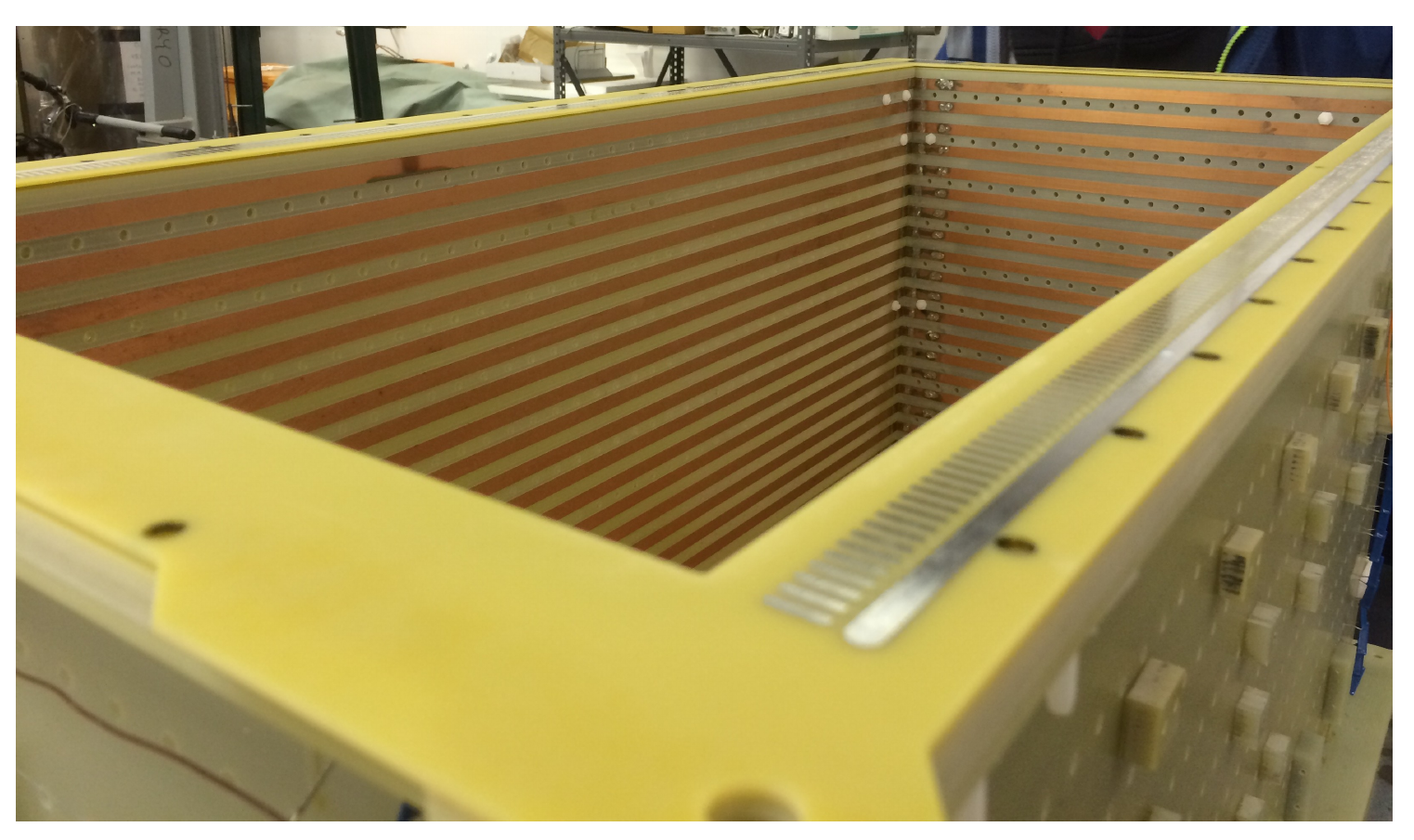}
\includegraphics[width=3.5in]{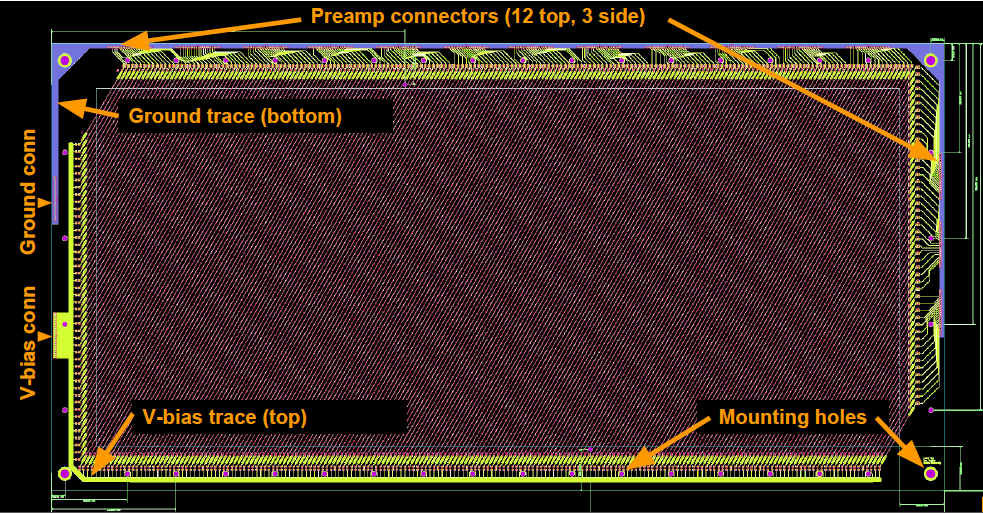}
%\vspace{-4.3cm}
\caption{ {\scriptsize \sf Detail of the LArIAT LArTPC field cage structure and PC board schematic of the first plane to see drifting electrons.}}
\label{fig:fieldcage-wireplane}
\end{centering}
\end{figure}
%%%%%%%%%%%%%%%%%%%%%%%%%%%%%%%%%%%%%%%%%%%%%
The field cage of the existing ArgoNeuT TPC shown in Fig.~\ref{fig:fieldcage-wireplane} is refurbished for the test-beam data run, with modifications to accommodate the installation of the cold electronics boards. The wire planes of the TPC are replaced by  a new set of planes.  The wire plane geometry is unchanged, but the new printed circuit design shown in Fig.~\ref{fig:fieldcage-wireplane} for the wire connection to the electronics has been adopted.
The wire spacing, or pitch, is 4~mm in all planes. The first plane to see the drifting electrons contains 225 parallel equal length wires, vertically oriented with respect to the ground and perpendicular to the beam axis.  This plane is not instrumented but serves to shape the electric field near the wire-plane and to shield the outer, instrumented planes against induction signals from the ionization charges while they are drifting through the LArTPC volume. The second, ``induction", plane consists of 240 wires oriented at $\mathrm{+60^{\circ}}$ relative to the beam axis.   The third and final, ``collection", plane is made up of 240 wires oriented at $\mathrm{-60^{\circ}}$ relative to the beam axis. The wires of the induction and collection planes are of varying lengths. 
%%%%%%%%%%% Figure 14 %%%%%%%%%%%%%%%%%%%%%%%%%%%%%%
\begin{figure}[h!]
\begin{centering}
\includegraphics[width=3.5in]{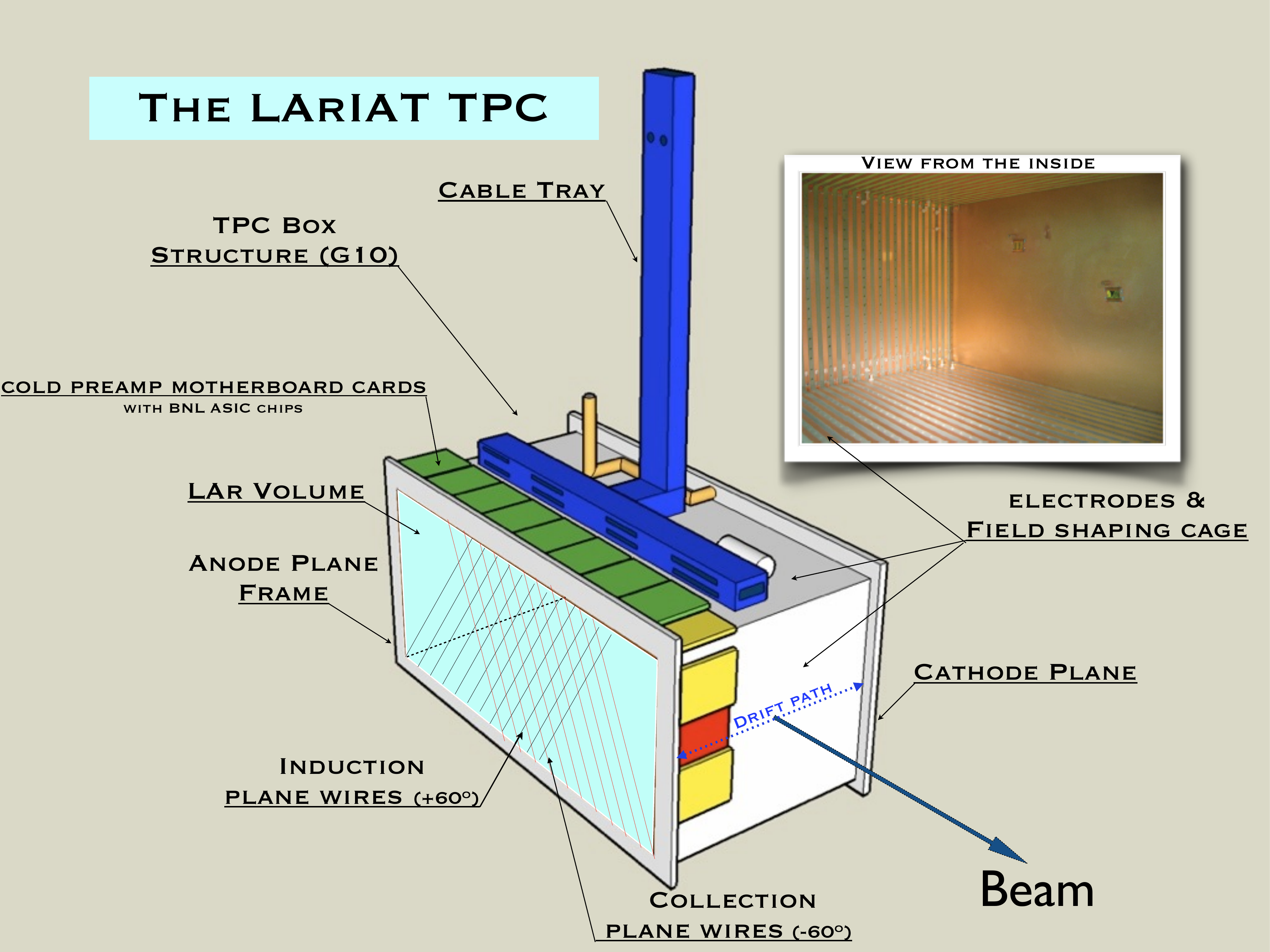}
%\vspace{-4.3cm}
\caption{{\scriptsize \sf Pictorial view of the LArTPC mechanics. Details of the anodic structure with the $\pm 60^\circ$ inclined wire-planes are indicated. The new cold electronics boards installed onto the TPC structure are also schematically indicated. In the insert, a picture of the inside of the LArTPC volume showing the cathodic plane and the copper strips of the field shaping cage.}}
\label{fig:TPC_schematic}
\end{centering}
\end{figure}
%%%%%%%%%%%%%%%%%%%%%%%%%%%%%%%%%%%%%%%%%%%%%

The TPC active volume is  $47~w\times 40~h\times 90~l~$cm$^{3}$, corresponding to a volume of 170~liters of liquid argon. This volume is delimited by a rectangular box structure acting as field shaping system of the TPC. The structure is composed of G10 manufactured by PCB technique with copper strips 1~cm wide spaced at 1~cm intervals, forming 23 rectangular rings all the way across the TPC as seen in Fig.~\ref{fig:TPC_schematic}. The cathode is a G10 plain sheet with copper metallization on the inner surface opposite the anode wire planes. The electric field is uniform over the entire TPC drift volume with a nominal value of 500 V/cm from the cathode to the anode planes, with a maximum drift length of $\ell_d=47$ cm. A new set of tensioning bars are installed to bring the wires to the desired tension without  interference with the PMTs of the light-colleciton system positioned behind the wire planes. A new cable-tray attached to the top of the TPC acts as a conduit to shield and route the readout cables to the top flange of the cryostat. The cables will also be replaced with ones insulated with a material of reduced water content.

\subsection{TPC Cold Electronics}
The new cold electronics read-out system for LArIAT consists of the following key components, shown in Fig.~\ref{fig:cold-elec}[Top]: 
%%%%%%%%%%% Figure 15 %%%%%%%%%%%%%%%%%%%%%%%%%%%%%%
\begin{figure}[h!]
\centering
\includegraphics[width=3.5in]{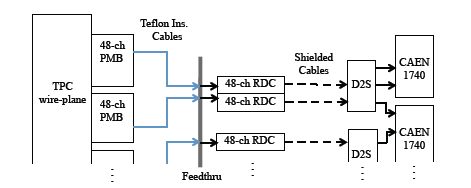}
\includegraphics[width=3.2in]{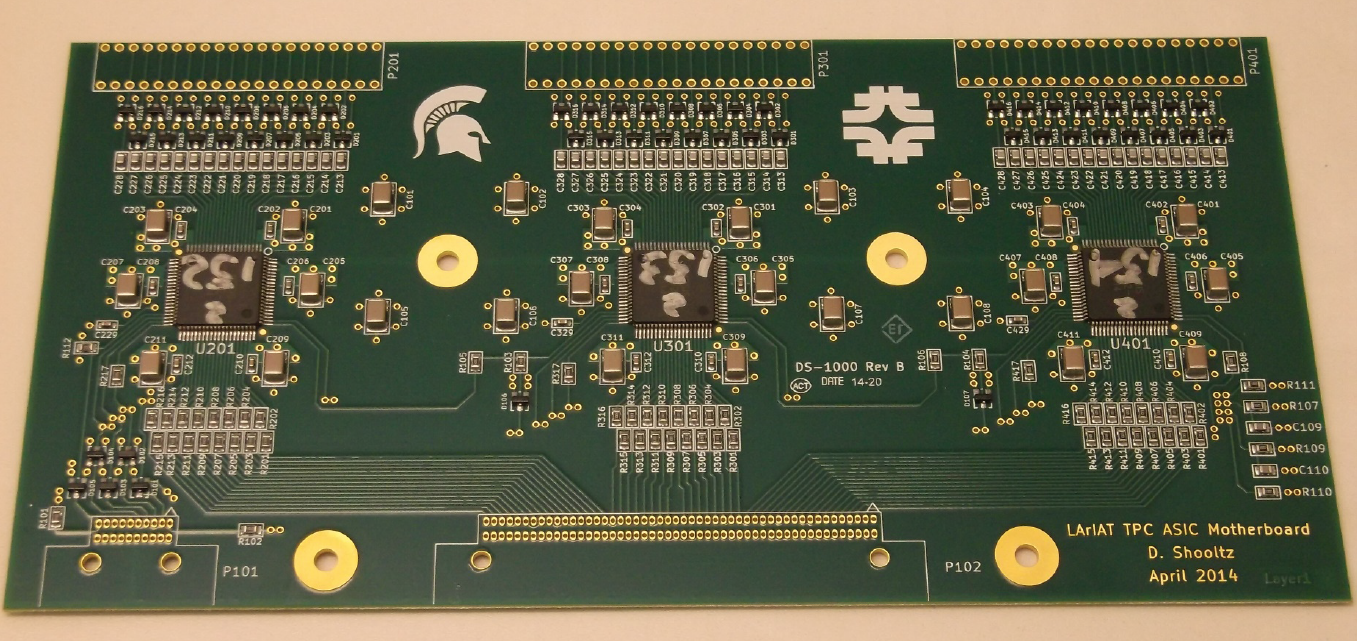}
\caption{{\scriptsize \sf A schematic representation of the read-out electronics layout and cold preamp motherboards with ASIC chips.} }
\label{fig:cold-elec} 
\end{figure}
%%%%%%%%%%%%%%%%%%%%%%%%%%%%%%%%%%%%%%%%%%%%
$i)$ new TPC-mounted cold preamp motherboards with ASIC chips designed by Brookhaven National Lab and used by the MicroBooNE experiment, $ii)$ Teflon insulated cables, with minimal water vapor outgassing used for the signal feed-through at the inner side,  $iii)$ warm Receiver-Driver Cards converting the signals from the LArTPC to differential signals, $iv)$ shielded cables that run to a location near the digitizers, $v)$ differential to single-ended signal cards, $vi)$ cables to the digitizers. For faster readout a new 64-channel CAEN 1740 digitizer is adopted. \\
A picture of a cold preamp motherboard from the recent production at MSU is shown in Fig.~\ref{fig:cold-elec}[Bottom].

%%%%%%%%%%% Figure 16 %%%%%%%%%%%%%%%%%%%%%%%%%%%%%%
\begin{figure}[h!]
\centering
\includegraphics[width=2.9in]{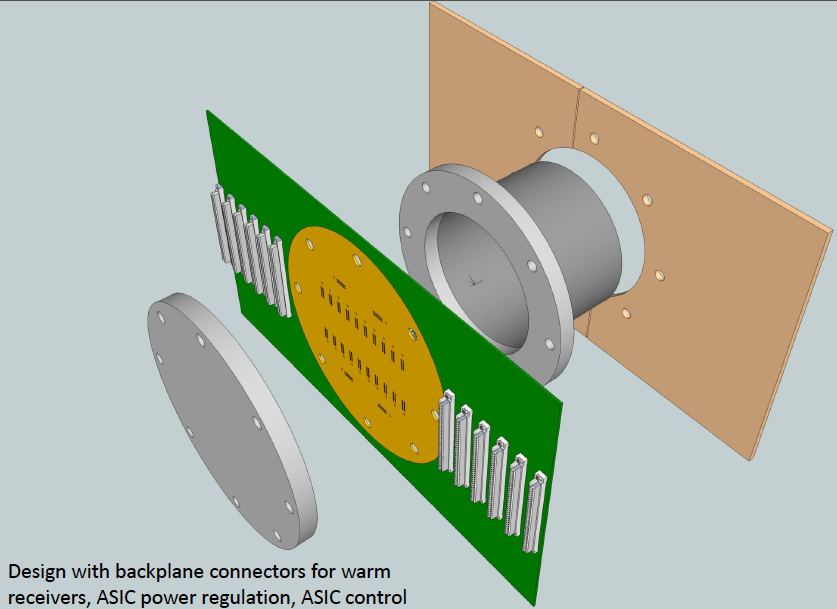}
\caption{{\scriptsize \sf  Design overview of the signal feed-through, showing the PCB card, the cryostat flange and the rear plate of the shielding box for the warm receivers.} }
\label{fig:new-FT} 
\end{figure}
%%%%%%%%%%%%%%%%%%%%%%%%%%%%%%%%%%%%%%%%%%%%
A multi-connector  high-vacuum custom feed-through for the wire signals has also been designed at MSU, Fig.~\ref{fig:new-FT}.  The signal cables plug from the inside and the warm receiver and control cards plug on the outside.

\subsection{Scintillation light read-out system}
LArIAT will have a high efficiency light readout system derived from solutions developed for Dark Matter liquid argon detectors. The aim is to extend the use of the scintillation light read-out  beyond the traditional triggering purposes of the liquid argon neutrino detectors to calorimetric energy reconstruction.  This improvement is achieved through measuring the fraction of energy deposited into scintillation light and using pulse shape discrimination for PID. The key element of the system is an array of two HQE photo-multiplier tubes (PMTs) for cryogenic applications, supplemented by the addition of two silicon photo-multiplier (SiPM) detectors, all deployed in liquid argon and mounted behind the wire planes of the TPC. Because LArIAT is re-using the ArgoNeuT TPC and cryostat some constraints are enforced on the system.  For example, the PMTs must fit into the side flange of the cryostat, which has been modified for this purpose.
%%%%%%%%%%% Figure 17 %%%%%%%%%%%%%%%%%%%%%%%%%%%%%%
\begin{figure}[h]
\begin{centering}
 \includegraphics[width=3.5in]{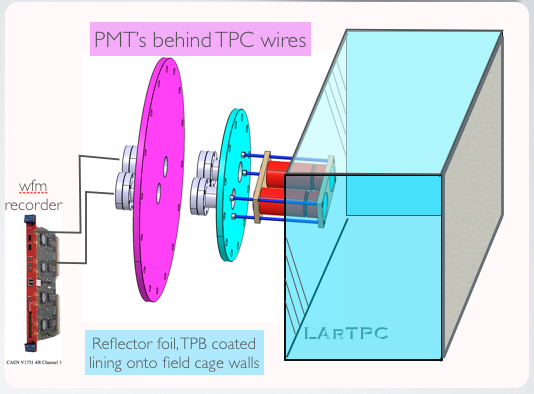}
 \caption{ {\scriptsize \sf The PMT array for scintillation light read-out on the cryostat side port and the waveform recorder for scintillation light read-out.}  }
\label{PMT_mod}
\end{centering}
\end{figure}
%%%%%%%%%%%%%%%%%%%%%%%%%%%%%%%%%%%%%%%%%%%%

Liquid argon scintillation light is emitted at 128 nm in the VUV range, outside the sensitivity of the PMTs. Wavelength-shifting (WLS) into the visible-blue is then necessary. The optical system coupled to to the PMT and SiPM array thus consists of a WLS-coated dielectric substrate layer lining on the inner surfaces of the field shaping system delimiting the active volume of the TPC. The WLS coating is made of a thin TetraPhenylBoutadiene (TPB) film deposited by vacuum evaporation. The dielectric layer is made of VIKUITI foils, a ultra-high specular reflectivity polymeric substrate. The scintillation VUV photons are wavelength-shifted into visible photons when hitting the TPB and the latter are reflected from the mirror surfaces beneath to be ultimately collected by the PMTs. The signals from the PMTs are digitized using a fast ADC, CAEN model V1751, and then stored.  The reflector coverage increases the uniformity of light collection and the fast  signal digitization enables to differentiate fast and slow light components. The setup is schematically shown in Fig.~\ref{PMT_mod}. This light readout system collects much more light than typical liquid argon neutrino experiments. The estimate from simulations points to about 50 photo-electrons/MeV at zero field, a value substantially higher than the estimated light yield of current liquid argon neutrino detectors  which is in the range of 1 photo-electron/MeV for ICARUS and  $\sim$2 photo-electrons/MeV for MicroBooNE.  The LBNE proposed system would only collect $\sim 0.2-0.3$ photo-electrons/MeV. The LArIAT expected yield has been recently confirmed through a direct measurement on a small scale prototype at University of Chicago - Apr.'14, \cite{UoC}.  

This design will allow in situ gain calibration using single photo-electrons from the tail of the signal while simultaneously allowing for event by event calorimetric reconstruction. This technique will be especially powerful for lower energy events that are usually close to the threshold for standard LArTPC detectors. An improvement of the calorimetric energy resolution up to a factor two is expected when the scintillation light signal is combined with TPC charge signal \cite{szydagis}. Advancements or new methods for particle identification are also anticipated \cite{Sorel}.

\subsection{Beam Detectors}
A system of beam counters is being aligned along the LArIAT tertiary beam line for beam particle selection and tracking. The layout is schematically shown in Fig.~\ref{fig:beam-counters}. It consists of a TOF plastic scintillator system (T1-3), a series of multi-wire proportional chambers (MWPC1-4), a Cherenkov counter and a scintillator paddle veto system (V1-2) located upstream of the TPC, for beam halo, and downstream for exiting particles.
%%%%%%%%%%% Figure 18 %%%%%%%%%%%%%%%%%%%%%%%%%%%%%%
\begin{figure}[h!]
\centering
\includegraphics[width=3.5in]{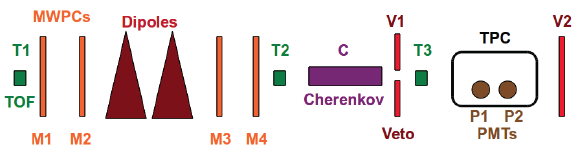}
\caption{{\scriptsize \sf A schematic layout of the LArIAT beam counters.} }
\label{fig:beam-counters} 
\end{figure}
%%%%%%%%%%%%%%%%%%%%%%%%%%%%%%%%%%%%%%%%%%%%
Another system made of two staggered arrays of plastic scintillator paddles in time coincidence for $\sim$horizontal cosmic muon selection between beam spills is also being positioned. The collection of slanted muon tracks diagonally crossing the TPC volume from cathode to anode wire plane will allow continuous monitoring of the liquid argon purity through measurement of the ionization electron lifetime.

\subsection{Trigger and DAQ}
The LArIAT DAQ and trigger system consists of the modular electronics and computer systems which initiate detector readout, do analog to digital conversion, interface and synchronize with the accelerator system, store raw data on disk, transfer data to archival storage and conduct first pass data quality monitoring. The system, shown in Fig.\ref{fig:DAQ}, is a mixture of commercial VME electronics from CAEN, a VME time to digital convertor for the beamline time of flight system, and a readout system for beamline wire chambers. The latter two components were produced by Fermilab. There are also a pair of NIM bins (not shown) with level translators, fan-outs, and gate generators which are needed to interface with the accelerator complex and the beamline counters. The system is controlled by a rack mounted computer via a PCIe optical interface to the digitizer modules in one VME crate and also to the controller board in a second crate. The DAQ resides on a private network accessible via a gateway machine. Data are written to a partition shared between the gateway and the DAQ machine, permitting online monitoring by processes on the gateway or on PCs in the external network. 
%%%%%%%%%%% Figure 19 %%%%%%%%%%%%%%%%%%%%%%%%%%%%%%
\begin{figure}[h!]
\centering
\includegraphics[width=3.3in]{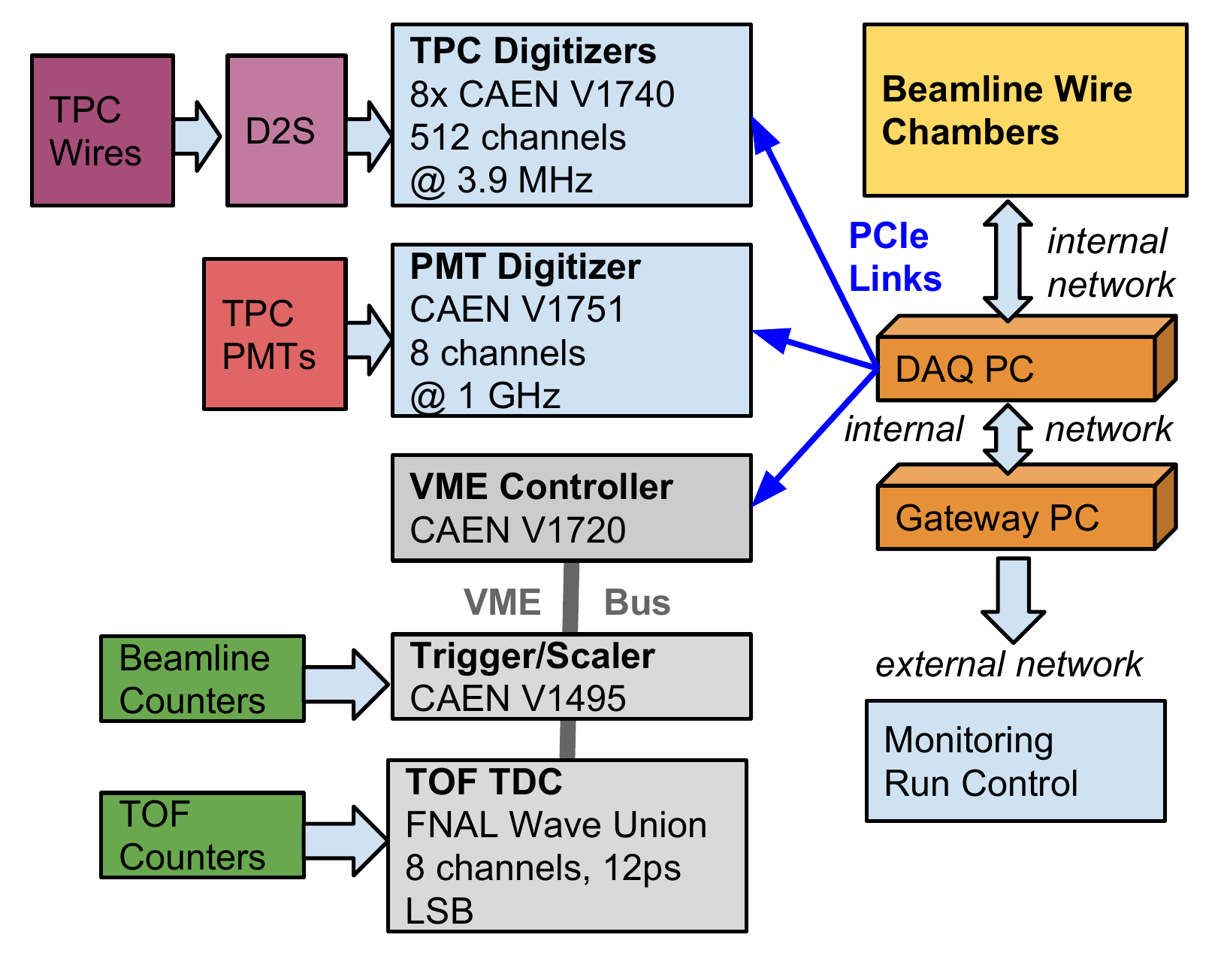}
\caption{ {\scriptsize \sf  Schematic of DAQ and Trigger System.} }
\label{fig:DAQ}
\end{figure}
%%%%%%%%%%%%%%%%%%%%%%%%%%%%%%%%%%%%%%%%%%%%

The FNAL accelerator complex sends one 4sec long beam spill to the LArIAT experimental area every $\sim$60sec supercycle. Just before the spill an accelerator signal is used to derive a clock reset which is fanned out to each of the CAEN waveform digitizer boards, the TOF unit, and the trigger. The trigger module takes digital inputs from each of the beamline TOF counters, halo veto counters in front of the detector, punchthrough counters at the rear, a Cherenkov detector, and the logical OR of the wires in each wire chamber. It has custom firmware which looks for user programmable patterns corresponding to ?good? events. A good pattern results in a fast trigger output and starts a clock. The clock counts for the maximum drift time in the TPC (roughly 330 microseconds) and the board then outputs a delayed trigger. The fast output is used to trigger the PMT digitizers while the slow output is used to record the last 330 microseconds from the TPC digitizers. The trigger contains logic to veto on ?bad? patterns and inhibit TPC readout if a second good or bad pattern occurs during the drift time. The TOF TDC and wire chambers read out continuously, and with zero dead-time, during the spill.  At the end of the spill the DAQ creates a single record consisting of the waveforms from each TPC wire and PMT channel for each triggered event, as well as data from the TOF and wire chambers, and a summary from the trigger board. The data from all boards is timestamped to permit offline event building. 

The DAQ is implemented in C/C++ and runs as a standard UNIX executable. Configuration and driver libraries have been completed for each of the modules and the full system can run in the beam data-taking mode described above. A more sophisticated version of the DAQ, implemented in the Fermilab ?artdaq? framework, is under parallel development. It completely reuses the configuration and driver libraries but has the potential for more advanced run control, logging, and data monitoring capabilities.

\subsection{Cooling and Purifications system}
The ArgoNeuT cooling and purification system relied on purification of the gaseous argon only. The flow rate was low, providing a full volume exchange every 7-8~days. For faster, more stable, and reliable operation, a new dedicated skid for combined gaseous and liquid argon purification has been specified by the Fermilab cryo-engineering team and designed by a vendor.
%%%%%%%%%%% Figure 20 %%%%%%%%%%%%%%%%%%%%%%%%%%%%%%
\begin{figure}[h!]
\centering
\includegraphics[width=3.5in]{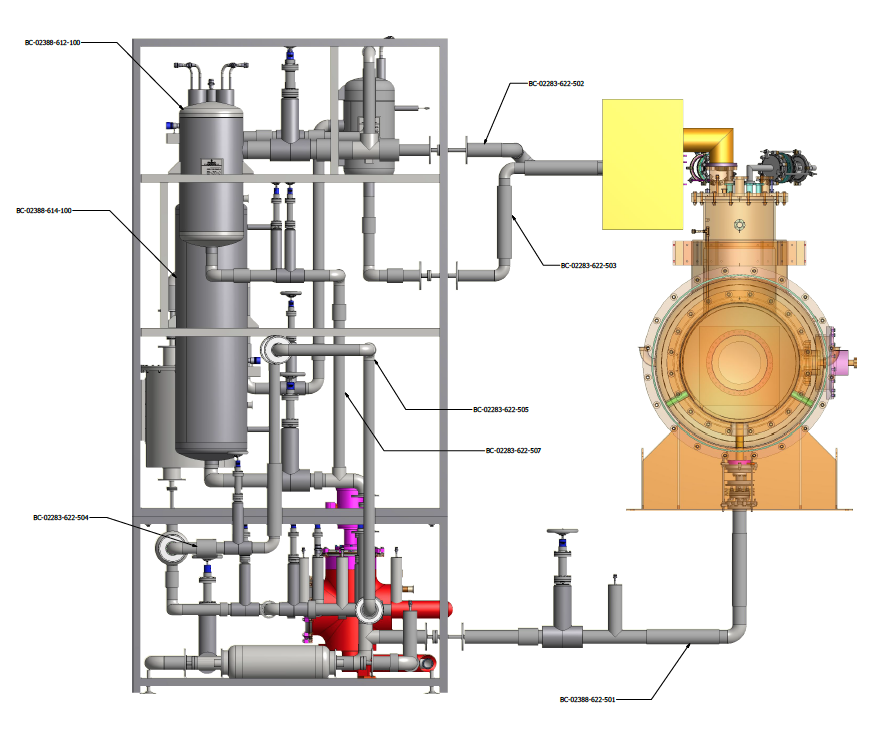}
\includegraphics[width=3.5in]{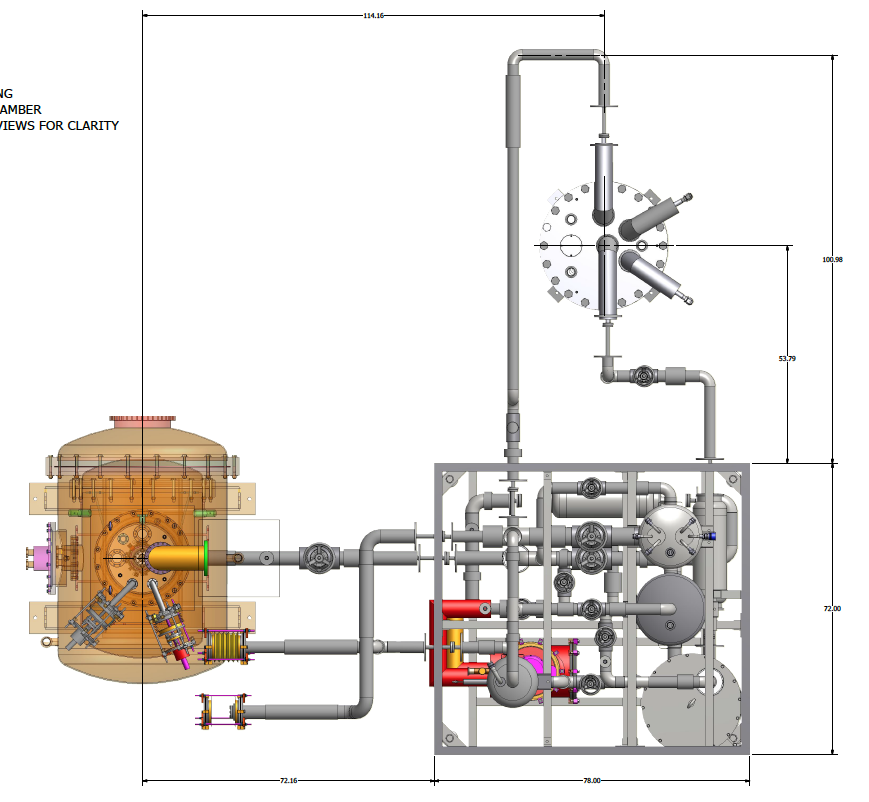}
\caption{ {\scriptsize \sf  Schematic views of the cryo-purification system: Front view [top], Top view [bottom]. The ArgoNeuT cryostat, the liquid argon pump and filter for oxygen and water removal are part of the skid.} }
\label{fig:uB-pump-filter}
\end{figure}
%%%%%%%%%%%%%%%%%%%%%%%%%%%%%%%%%%%%%%%%%%%%
The new system is composed of a forced-recirculation network of vacuum-jacketed and super-insulated transfer lines, where boil-off gaseous argon is extracted upstream of a liquid nitrogen condenser outside of the cryostat and liquid argon from main cryostat is extracted from the outlet at its bottom. The recondensed argon and liquid pulled from the bulk volume mix and via gravity reach the head of a centrifugal cryogenic pump housed in a vacuum-insulated vessel. A downstream filtering system skid, made of a large vacuum-jacketed cartridge filled with a bed of molecular sieve and activated copper material, removes moisture and oxygen from the argon. The purified liquid is
returned to the main cryostat through the last portion of the transfer line network, shown in Fig.~\ref{fig:uB-pump-filter}.  The pump skid and the filter skid are clone-copies of the systems developed and designed for LAPD~\cite{LAPD}.  The flow capability of the pump skid allows recycling the liquid argon in the LArIAT vessel in less than two hours. The circulation system is configured to ensure that the pump net positive suction will be adjustable  for the different operations. The pump is known to have much larger capacity than required for Phase-1 of the LArIAT program and the high flow is accommodated by splitting part of it in a sub-loop circuit. The excess energy is removed through the high cooling power condenser unit.

\section{LArIAT Phase-2: outlook}
Because of the much larger size of the LArTPC in Phase-2, LArIAT will be able to repeat the measurements of Phase-1 with much better containment.  It will also extend its physics reach to more fully study hadronic interactions in liquid argon.  Figure~\ref{fig:containment} shows the fraction of charged pion energy contained in liquid argon as a function of longitudinal and transverse distance from the pion starting position.  This figure shows that a detector whose transverse dimension to the beam is at least 1~m from the beam spot and whose longitudinal size is at least 3~m will contain 90\% of the total pion energy on average.  The same study showed that at least 20\% of those pions will have 95\% of their energy contained in that volume.

%%%%%%%%%%% Figure 21 %%%%%%%%%%%%%%%%%%%%%%%%%%%%%%
\begin{figure}[h!]
\centering
\includegraphics[width=3.in]{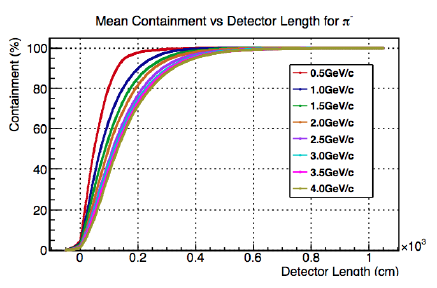}
\includegraphics[width=3.in]{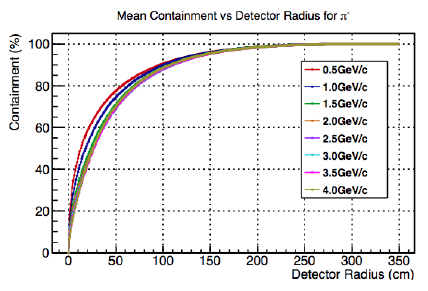}
\caption{\scriptsize \sf Fraction of charged pion energy contained in liquid argon as a function of longitudinal [Top] and transverse [Bot] distance from its starting position.}
\label{fig:containment}
\end{figure}
%%%%%%%%%%%%%%%%%%%%%%%%%%%%%%%%%%%%%%%%%%%%
Phase-2 will provide detailed studies of the calorimetry possible with a LArTPC through three methods:
\begin{itemize}
\item  ``Traditional'' calorimetry where the electrons ionized by charged particles induce signals in the wire planes that are directly related to the energy of the primary particle. For hadrons, this ionization energy accounts only for a portion of the incident energy with the rest being lost to binding energy as multiple low energy protons and neutrons are liberated from nuclei participating in the cascade. Quantifying the fraction of this ``invisible" energy to the total incoming energy of the hadron is a primary deliverable of a test beam experiment.
\item  ``Scintillation" calorimetry where the produced scintillation light is detected by photosensors placed inside the cryostat. The energy deposited in the detector is divided between the scintillation and ionization signals and understanding how to combine those signals will yield an optimal measurement of particle energies.  
\item ``Topological" calorimetry where traditional calorimetry can be augmented with information about the event topology, such as the kinematics information and particle species of secondary tracks. 
\end{itemize}
Placing a LArTPC in a test beam allows one to develop each of these methods using particles of known energy and quantify the resolution of each.  A combination of all methods will likely yield the most precise energy measurements for the observed neutrino interactions and validating these methods in a test beam will also be very important for LBNE.  The containment offered by the Phase-2 cryostat will be very important for providing precise calibration of the calorimetric energy response of the detector to hadrons.  As seen in Fig.~\ref{fig:mom-spectra}, charged pions compose a significant fraction of the particles emerging from charged-current neutrino interactions. 

In addition, Phase-2 also offers the opportunity to study cosmic ray backgrounds and their impact on beam related activity.  It will operate in a high-rate environment for cosmic rays and the studies of how frequently those cosmic rays interfere with the beam particles in reconstruction will be very useful. 

\section{Conclusion}
A comprehensive characterization of LArTPC performance is now considered of great interest and impact for the development of the medium- and long-term Intensity Frontier program in the U.S.~\cite{LArTPC-R&D}.  Specifically, this characterization is required for particles such as $\gamma,~e^\pm,~\mu^\pm,~\pi^{0,\pm},~k^\pm,~p,~\bar p$ in the range of energies from a few hundred MeV to 2.0~GeV.  This range of particles and energies is relevant to the forthcoming MicroBooNE experiment, and to the correlated SBN program at Fermilab, with the LAr1-ND and the ICARUS T600 detector in the near and far location respectively.  The particles and energies are also relevant LBN program for precise neutrino oscillation physics and proton decay searches.  The beam in MC7 is the ideal place to perform this type of study, providing not only a range of selectable known energies, but also a complete set of selectable types of different particles in both polarities.  The test beam also provides a controlled environment in which to tune simulations and to validate the off-line software tools for PID, calorimetry, and event reconstruction without relying solely on simulation. {\sf LArIAT} Phase-1 will repurpose the fine-grained LArTPC from ArgoNeuT for an extended physics run with low momenta charged particles.

}
%----------------------------------------------------------------------------------------
%	REFERENCE LIST
%----------------------------------------------------------------------------------------

\clearpage

\end{document}